\documentclass[
	journal,
	letterpaper,
	twoside,
	twocolumn
]{IEEEtran}

\usepackage[amssymb]{SIunits}
\usepackage[cmex10]{amsmath} 
\usepackage{amssymb}
\usepackage[latin1]{inputenc}
\usepackage{xcolor}
\usepackage{cite}
\usepackage[pdftex]{graphicx}
\usepackage[caption=false, font=footnotesize]{subfig}
\usepackage{url}

\newenvironment{DIFnomarkup}{}{}

\usepackage[
	shortcuts,
	nonumberlist,	
	acronym, 
	toc, 
	hyperfirst=true,
	section=section, 
	order=letter,
	numberedsection=nolabel 
]{glossaries}

\newglossary[slg]{los}{syi}{syg}{List of Symbols} 
\makeglossaries 
\glsdisablehyper 
\newglossarystyle{mylist}{
	\glossarystyle{list} 
}

\newglossarystyle{mysymbolstyle}{

	\renewcommand*{\glsgroupheading}[1]{}%
}

\newglossarystyle{myacronymstyle}{
  \glossarystyle{mysymbolstyle} %

}
\newcommand{\NewS}[5][\newcommand]{
	\newglossaryentry{symb:#2}{
		name=\ensuremath{#3},
		description={\nopostdesc #4}, 
		sort=sym#5,
		type=los,
	}
	\glsadd{symb:#2}
	\expandafter\def\csname #2\endcsname{\ensuremath{#3}} 
}

\newcommand{\NewC}[5][\glsadd]{
	\newglossaryentry{symb:#2}{
		name=\ensuremath{#3},
		description={\nopostdesc #4}, 
		sort=sym#5,
		type=los,
	}
	\expandafter\def\csname #2\endcsname{\ensuremath{#3}} 
}

\newcommand{\NewF}[6][\newcommand]{
	\newglossaryentry{symb:#2}{
		name=\ensuremath{#3\cdot#4},
		description={\nopostdesc #5}, 
		sort=sym#6,
		type=los,
	}
	\expandafter\def\csname #2\endcsname ##1{\ensuremath{#3{##1}#4}} 
}
\NewS{TXR}{R_0}{transmitted radius}{rx}
\NewS{txR}{r_0}{transmitted radius}{rx}

\NewS{TXP}{\Theta_0}{transmitted phase}{tx}
\NewS{txP}{\theta_0}{transmitted phase}{tx}

\NewS{RXR}{R}{received radius}{ry}
\NewS{rxR}{r}{received radius}{ry}

\NewS{RXP}{\Theta}{received phase}{ty}
\NewS{rxP}{\theta}{received phase}{ty}

\NewS{TXRdet}{\hat{R}_0}{transmitted radius}{rx}

\NewS{estX}{\hat{X}}{estimated symbol}{xh}
\NewS{esttxR}{\hat{R}_x}{estimated transmitted radius}{rxe}
\NewS{esttxP}{\hat{\Theta}_x}{estimated transmitted phase}{txe}

\NewS{NLPN}{\Phi_{\text{NL}}}{nonlinear phase noise}{phi}
\NewS{Len}{{L}}{total fiber length}{L}
\NewS{Power}{{P}}{input power}{L}

\newcommand{\ri}{r_k}
\newcommand{\rii}{r_{k+1}}

\newcommand{\normr}[1]{\tilde{r}_{#1}}
\newcommand{\normmu}[1]{\tilde{\mu}_{#1}}

\NewS{X}{X}{transmitted symbol}{x}
\NewS{Y}{Y}{received symbol}{y}
\NewS{setX}{\mathcal{X}}{signal constellation}{xs}
\NewS{setAPSK}{\setX_{\text{APSK}}}{APSK signal set}{xsapsk}

\NewS{rad}{r}{radius}{r}
\NewS{ppr}{l}{points per ring}{l}
\NewS{phaseoffset}{\phi}{phase offset per ring}{p}
\NewS{NumRings}{N}{number of rings in the APSK constelllation}{N}
\NewS{RD}{\bold{r}}{radii distribution}{rr}

\NewS{NumSpans}{K}{number of spans}{k}

\NewS{SEP}{\text{SEP}}{symbol error probability}{ty}

\NewS{labeling}{\mathbb{L}}{labeling}{l}
\NewS{BRGC}{\mathbb{G}}{binary reflected gray code}{gc}
\NewS{ODP}{\otimes}{ordered direct product}{odp}

\NewS{Reg}{\mathcal{R}}{compact bounding region}{reg}

\NewS{imag}{\jmath}{imaginary unit}{j} 

\NewC{Code}{\mathcal{C}}{code}{C:code}
\NewC{vecI}{\boldsymbol{I}}{identity matrix}{i}
\NewC{vecZero}{\boldsymbol{0}}{all-zero vector}{0z}
\NewC{xor}{\oplus}{exclusive or}{0excl}
\NewC{SNR}{\text{SNR}}{signal to noise ratio}{s}
\NewC{mod}{\;\operatorname{mod}\,}{modulo operation}{m}	
\NewC{cconv}{\circledast}{circular convolution}{0}	

\NewC{natural}{\mathbb{N}}{set of natural numbers}{n} 
\NewC{real}{\mathbb{R}}{set of real numbers}{rs}
\NewC{rational}{\mathbb{Q}}{set of rational numbers}{q}
\NewC{integer}{\mathbb{Z}}{set of integer numbers}{zzzz}
\NewC{complex}{\mathbb{C}}{set of complex numbers}{c}

\NewC{Rn}{\mathbb{R}^n}{real Euclidean $n$-dimensional space}{rsn}
\NewC{GF}{\mathbb{F}_2}{Galois field of size two}{f}

\NewC{N}{\mathcal{N}}{normal distribution}{no}
\NewC{Norm}{\mathcal{N}}{normal distribution}{N}
\NewC{Unif}{\operatorname{Unif}}{uniform distribution}{uz}
\NewF{norm}{||}{||}{Euclidean norm}{0e}
\NewF{Qla}{Q_{\Lambda}(}{)}{nearest neighbor lattice quantizer}{quant}
\NewF{Qco}{Q_{\Code}(}{)}{binary quantizer with respect to a linear code $\Code$}{quant2}
\NewF{E}{\mathbb{E}\left[}{\right]}{expectancy operator}{e}
\NewF{bef}{\operatorname{H}(}{)}{binary entropy function}{he}
\NewF{Pr}{\operatorname{Pr}\left[}{\right]}{probability of an event}{pr}
\renewcommand{\E}[2][]{{\mathbb{E}#1\left[ #2 \right]}}			 

\newcommand{\bessel}[2][]{{\operatorname{I}_#1\left(#2\right)}}

\newcommand{\MarcumQfunc}[1]{Q_\text{1}\left(#1\right)}

\newcommand{\IE}{i.e., } 
\newcommand{\EG}{e.g., } 

\newcommand{\MYsectionname}{Sec. }
\newcommand{\define}{\triangleq}
\newcommand{\forwhich}{\; : \;}	
\newcommand{\vect}[1]{\boldsymbol{#1}}
\newcommand{\dB}[1]{\ensuremath{\unit{#1}{\deci\bel}}}
\newcommand{\dBm}[1]{\ensuremath{\unit{#1}{\deci\bel\metre}}}
\newcommand{\km}[1]{\ensuremath{\unit{#1}{\kilo\metre}}}

\newcommand{\D}{\mathrm{d}}

\renewcommand{\epsilon}{\varepsilon}
\renewcommand{\phi}{\varphi}

\makeatletter
\newcommand{\vast}{\bBigg@{5}}
\makeatother

\definecolor{shadecolor}{rgb}{0.97,0.97,0.97}%
\definecolor{framecolor}{rgb}{0,0,0}%

\newcommand{\abbr}[1]{{#1}}				

\makeatletter
\let\aclOLD=\acl
\renewcommand{\acl}[1]{%
  \begingroup    
  \let\@@underline=\relax
  \aclOLD{#1}%
  \endgroup
}
\makeatother

\newcommand{\NewA}[3]{
	\newacronym{#1}{#2}{#3}
}

\NewA{af}{AF}{\abbr{a}mplify-and-\abbr{f}orward}
\NewA{apsk}{APSK}{\abbr{a}mplitude \abbr{p}hase-\abbr{s}hift \abbr{k}eying}
\NewA{ask}{ASK}{\abbr{a}mplitude-\abbr{s}hift \abbr{k}eying}
\NewA{ase}{ASE}{\abbr{a}mplified \abbr{s}pontaneous \abbr{e}mission}
\NewA{awgn}{AWGN}{\abbr{a}dditive \abbr{w}hite \abbr{G}aussian \abbr{n}oise}
\NewA{bep}{BEP}{\abbr{b}it \abbr{e}rror \abbr{p}robability}
\NewA{ber}{BER}{\abbr{b}it \abbr{e}rror \abbr{r}ate}
\NewA{qap}{QAP}{\abbr{q}uadratic \abbr{a}assignment \abbr{p}roblem}
\NewA{bicm}{BICM}{\abbr{b}it-\abbr{i}nterleaved \abbr{c}oded \abbr{m}odulation}				
\NewA{bpsk}{BPSK}{\abbr{b}inary \abbr{p}hase-\abbr{s}hift \abbr{k}eying}				
\NewA{bsc}{BSC}{\abbr{b}inary \abbr{s}ymmetric \abbr{c}hannel}				
\NewA{brgc}{BRGC}{\abbr{b}inary \abbr{r}eflected \abbr{G}ray \abbr{c}ode}				
\NewA{cf}{CF}{\abbr{c}haracteristic \abbr{f}unction}
\NewA{csit}{CSIT}{\abbr{c}annnel \abbr{s}tate \abbr{i}nformation at the \abbr{transmitter}}
\NewA{csi}{CSI}{\abbr{c}annnel \abbr{s}tate \abbr{i}nformation}
\NewA{df}{DF}{\abbr{d}ecode-and-\abbr{f}orward}
\NewA{fd}{FD}{\abbr{f}ull-\abbr{d}uplex}
\NewA{fft}{FFT}{\abbr{f}ast \abbr{F}ourier \abbr{t}ransform }
\NewA{hd}{HD}{\abbr{h}alf-\abbr{d}uplex}
\NewA{iid}{IID}{\abbr{i}ndepend and \abbr{i}dentically \abbr{d}istributed }
\NewA{isi}{ISI}{\abbr{i}nter\abbr{s}ymbol \abbr{i}nterference }
\NewA{lb}{LB}{\abbr{l}ower \abbr{b}ound}
\NewA{map}{MAP}{\abbr{m}aximum \abbr{a} \abbr{p}osteriori}
\NewA{mf}{MF}{\abbr{m}odulo-and-\abbr{f}orward}
\NewA{mlan}{MLAN}{\abbr{m}odulo-\abbr{l}attice \abbr{a}dditive \abbr{n}oise}
\NewA{ml}{ML}{\abbr{m}aximum \abbr{l}ikelihood}
\NewA{mmse}{MMSE}{\abbr{m}inimum \abbr{m}ean \abbr{s}quare \abbr{e}rror}
\NewA{nlpc}{NLPC}{\abbr{n}onlinear \abbr{p}hase \abbr{c}ompensation}
\NewA{nlpn}{NLPN}{\abbr{n}onlinear \abbr{p}hase \abbr{n}oise}
\NewA{nc}{NC}{\abbr{n}etwork \abbr{c}oding}
\NewA{ofmd}{OFDM}{\abbr{o}rthogonal \abbr{f}requency-\abbr{d}ivision \abbr{m}ultiplexing }			
\NewA{dp}{DP}{\abbr{d}ual-\abbr{p}olarization}
\NewA{pam}{PAM}{\abbr{p}ulse \abbr{a}mplitude \abbr{m}odulation}
\NewA{pdf}{PDF}{\abbr{p}robability \abbr{d}ensity \abbr{f}unction}
\NewA{plnc}{PNC}{\abbr{p}hysical-layer \abbr{n}etwork \abbr{c}oding}
\NewA{psk}{PSK}{\abbr{p}hase-\abbr{s}hift \abbr{k}eying}
\NewA{pmd}{PMD}{\abbr{p}olarization \abbr{m}ode \abbr{d}ispersion}
\NewA{pdm}{PDM}{\abbr{p}olarization-\abbr{d}ivision \abbr{m}ultiplexing}
\NewA{qam}{QAM}{\abbr{q}uadrature \abbr{a}mplitude \abbr{m}odulation}
\NewA{sqp}{SQP}{\abbr{s}equential \abbr{q}uadratic \abbr{p}rogramming}
\NewA{rd}{RD}{\abbr{r}adii \abbr{d}istribution}
\NewA{sep}{SEP}{\abbr{s}ymbol \abbr{e}rror \abbr{p}robability}
\NewA{ser}{SER}{\abbr{s}ymbol \abbr{e}rror \abbr{r}ate}
\NewA{si}{SI}{\abbr{s}ide \abbr{i}nformation}
\NewA{sp}{SP}{\abbr{s}ingle-\abbr{p}olarization}
\NewA{snr}{SNR}{\abbr{s}ignal-to-(additive-)\abbr{n}oise \abbr{r}atio}
\NewA{snlse}{sNLSE}{\abbr{s}tochastic \abbr{n}onlinear \abbr{S}chr\"odinger \abbr{e}quation}
\NewA{stwrc}{sTRC}{\abbr{s}eparated \abbr{t}wo-way \abbr{r}elay \abbr{c}hannel}
\NewA{stwtrc}{sTTRC}{\abbr{s}eparated \abbr{t}wo-way \abbr{t}wo-\abbr{r}elay \abbr{c}hannel}
\NewA{ts}{TS}{\abbr{t}wo-\abbr{s}tage}
\NewA{twrc}{TRC}{\abbr{t}wo-way \abbr{r}elay \abbr{c}hannel}
\NewA{twtrc}{TTRC}{\abbr{t}wo-way \abbr{t}wo-\abbr{r}elay \abbr{c}hannel}
\NewA{wdm}{WDM}{\abbr{w}avelength \abbr{d}ivision \abbr{m}ultiplexing}

\NewA{dm}{DM}{dispersion-managed}

\NewA{spm}{SPM}{self-phase modulation}
\NewA{xpm}{XPM}{cross-phase modulation}
\NewA{fwm}{FWM}{four-wave-mixing}

\NewA{ixpm}{IXPM}{intrachannel cross-phase modulation}
\NewA{ifwm}{IFWM}{intrachannel four-wave-mixing}
\NewA{ssfm}{SSFM}{split-step Fourier method}

\begin{document}

\begin{DIFnomarkup}

\title{Design of APSK Constellations for Coherent Optical Channels with Nonlinear Phase Noise}

\author{
	\IEEEauthorblockN{
	Christian Häger, \emph{Student Member, IEEE},
	Alexandre Graell i Amat, \emph{Senior Member, IEEE},
	Alex Alvarado, \emph{Member, IEEE}, and
	Erik Agrell, \emph{Senior Member, IEEE}
	\thanks{Parts of this paper have been presented at the
	2012 IEEE Global Communication Conference (GLOBECOM), Anaheim, CA,
	Dec. 2012.}
	\thanks{This work was partially funded by the Swedish Agency for
	Innovation Systems (VINNOVA) under the P36604-1 MAGIC project,
	Sweden, and by 
	the European Community's Seventh's Framework Programme
	(FP7/2007-2013) under grant agreement No. 271986.  The calculations were performed 
	on resources provided by the Swedish National Infrastructure
	for Computing (SNIC) at C3SE.} 
	\thanks{C. Häger, A. Graell i Amat, and Erik Agrell are with the
	Department of Signals and Systems, Chalmers University of Technology, Gothenburg,
	Sweden (email: \{christian.haeger, alexandre.graell,
	agrell\}@chalmers.se). A. Alvarado is with the Department of 
	Engineering, University of Cambridge, UK (email:
	alex.alvarado@ieee.org).}
	}
}

\maketitle

\end{DIFnomarkup}

\begin{abstract}
	We study the design of \gls{apsk} constellations for a coherent
	fiber-optical communication system where \gls{nlpn} is the main
	system impairment. \gls{apsk} constellations can be regarded as a
	union of \gls{psk} signal sets with different amplitude levels. A
	practical \gls{ts} detection scheme is analyzed, which performs
	close to optimal detection for high enough input power.  We optimize
	\gls{apsk} constellations with 4, 8, and 16 points in terms of
	\gls{sep} under \gls{ts} detection for several combinations of input
	power and fiber length. 
	For 16 points, performance gains of \dB{3.2} can be
	achieved at a \gls{sep} of $10^{-2}$ compared to 16-QAM by choosing
	an optimized \gls{apsk} constellation. We also demonstrate that in
	the presence of severe nonlinear distortions, it may become
	beneficial to sacrifice a constellation point or an entire
	constellation ring to reduce the average \gls{sep}. Finally, we
	discuss the problem of selecting a good binary labeling for the
	found constellations. 
\end{abstract}

\glsresetall 

%
%

\section{Introduction}
\label{sec:introduction}

Fiber nonlinearities are considered to be one of the limiting factors
for achieving high data rates in coherent long-haul optical
transmission systems \cite{Djordjevic2010, Agrawal2006,
Ho2005}. Therefore, a good understanding of the influence of
nonlinearities on the system behavior is necessary in order to
increase data rates of future optical transmission systems.

The optimal design of a signal constellation, \IE placing $M$ points
in the complex plane such that the \gls{sep} is minimized under an
average or peak power constraint, can be considered a classical
problem in communication theory \cite[Ch. 1]{Hanzo2000}. The problem
was addressed for example by Foschini \emph{et al.} in the early 70s
for the \gls{awgn} channel with \cite{Foschini1973} and without
\cite{Foschini1974} considering a random phase jitter.  However, only
little is known about the influence of fiber nonlinearities on the
optimal signal set. In this paper, we consider signal constellation
design for a nonlinear fiber-optical channel model assuming
single-channel transmission, hence neglecting interchannel
impairments. We focus on a specific class of constellations called
\gls{apsk}, which can be defined as the union of \gls{psk} signal sets
with different amplitude levels. This choice is motivated by the fact
that these constellations have long been recognized to be robust
formats to cope with nonlinear amplifier distortions prevalent in
satellite communication systems, see, \EG \cite{Thomas1974,
Biglieri1984, DeGaudenzi2006a, Sung 2009}, \cite[pp.
27--28]{Hallo2009} and references therein. 

%
%
%

The input--output relationship of the fiber-optical baseband channel
is described implicitly by the \gls{snlse} \cite[Ch.
2]{Agrawal2010}. It is well recognized that this type of
channel model does not lend itself to an easy solution for various
communication theoretic problems \cite{Gobel2010, Essiambre2010}. We
therefore consider a simplified, dispersionless channel model
which follows from the \gls{snlse} by neglecting the dispersive term
and captures the interaction of Kerr-nonlinearities with the signal
itself and the inline \gls{ase} noise, giving rise to \gls{nlpn}
\cite{Gordon1990,Mecozzi1994}. A discrete channel is obtained
from the waveform channel on a per-sample basis (assuming ideal
carrier and timing recovery) \cite[Sec. III]{Yousefi2011a}. This model
has been previously considered by several authors in the literature
and different methods have been applied to derive the joint \gls{pdf}
of the received amplitude and phase \cite{Mecozzi1994,
Turitsyn2003, Ho2005, Yousefi2011a}. Since all these
derivations neglect dispersion, the resulting \gls{pdf} should serve
as a useful approximation for \gls{dm} optical links, provided that
the local accumulated dispersion is sufficiently low \cite[p.
160]{Ho2005}, \cite{Ip2010a}. However, if the interaction between
dispersion and nonlinearities becomes too strong, the channel model is
likely to diverge from the one assumed here.\footnote{As an extreme
case, for dispersion-uncompensated links, it was found that the
channel is well-modeled by a Gaussian \gls{pdf} \cite{Carena2012,
Beygi2012}.} We point out that several studies have addressed the
influence of dispersion on the variance of \gls{nlpn} in the context
of \gls{dm} links using linearization techniques \cite{Green2003,
Kumar2005, Ho2006, Demir2007, Kumar2009a}.  In \cite{Bononi2010} a
comprehensive study on quantifying the parameter space where nonlinear
signal-noise interactions including \gls{nlpn} are dominant
impairments for different modulation formats was presented. A brief
discussion on the applicability of the assumed channel model in the
context of \gls{dm} links is also provided in \cite{Bononi2010}. An
extensive literature review on the topic of \gls{nlpn} is included in
\cite{Demir2007}.

Signal constellation design and detection assuming the same channel as
here has been studied previously in \cite{Lau2007d, Beygi2011,
Ekanayake2013}. In \cite{Lau2007d}, the authors applied several
predistortion and postcompensation techniques in combination with
minimum-distance detection for \gls{qam} to mitigate the effect of
\gls{nlpn}. They also proposed a \gls{ts} detector consisting of a
radius detector, an amplitude-dependent phase rotation, and a phase
detector. Moreover, parameter optimization was performed with respect
to four 4-point, custom constellations under \gls{ml} detection. In
\cite{Beygi2011}, the \gls{ts} detector was used to optimize the radii
of four 16-point constellations for two power regimes. It was shown
that the optimal radii highly depend on the transmit power.  In
\cite{Ekanayake2013}, the \gls{sep} of $M$-PSK was studied assuming a
minimum-distance detector. In \cite{Essiambre2010}, a capacity
analysis is presented for fiber-optical channels. The authors use
bivariate Gaussian \gls{pdf}s to represent the discrete-time channel
where the covariance matrices are obtained through extensive numerical
simulation. Continuous-input ring constellations are used to exploit
the assumed rotational invariance of the channel and subsequently find
lower bounds on the maximum achievable information rates. For the same
channel model, in \cite{Freckmann2009} the occupancy frequency and
spacing of the ring constellation were optimized. Related work was
presented in \cite{Zhang2011}, where the channel output \gls{pdf} is
approximated through numerically obtained histograms and optimized
ring constellations are found. Discrete constellations are then
obtained through quantization. A similar quantization technique was
applied in \cite{Djordjevic2012}. 



In this paper, we first analyze the (suboptimal) \gls{ts} detector
developed in \cite{Lau2007d}. We regard radius detection and
phase rotation as a separate processing block that yields a
postcompensated observation and we explain how to derive the
corresponding \gls{pdf} for constellations with multiple amplitude
levels. To the best of our knowledge, this \gls{pdf} has not been
previously derived, possibly due to the fact that the \gls{sep} under
\gls{ts} detection can be calculated with a simplified \gls{pdf}
\cite{Beygi2010}. The new \gls{pdf} is used to gain insights into the
performance behavior of the \gls{ts} detector compared to optimal
detection. We also show that this \gls{pdf} is necessary to accurately
calculate the average \gls{bep} of the constellation. 

We then find optimal \gls{apsk} constellations in terms of \gls{sep}
under \gls{ts} detection for a given input power and fiber length. In
contrast to \cite{Beygi2011}, we optimize the number of rings, the
number of points per ring, as well as the radii. For the case
$M = 4$, we choose identical system parameters as in
\cite{Lau2007d} and a comparison reveals that our approach
results in similar, sometimes better, constellations, with the
advantage of much less computational design complexity. This allows us
to extend the optimization to $M = 8$ and $M = 16$. For the latter
case, our results show that the widely employed 16-QAM constellation
has poor performance compared to the best found constellations over a
wide range of input powers for this channel model and 
detector. We also provide numerical support for the phenomenon of
\emph{sacrificial points} or \emph{satellite constellations}, which
arise in the context of constellation optimization in the presence of
very strong nonlinearities \cite{Agrell2012, Agrell2012a}. Our
findings show, somewhat counterintuitively, that it is sometimes
optimal to place a constellations point (or even an entire
constellation ring) far away from all other points in order to improve
the average performance of the constellation. 

Due to the separation of a hard-decision symbol detector and
subsequent error correction in state-of-the-art fiber-optical
communication systems, the uncoded \gls{bep} is an important figure of
merit. Therefore, we also address the problem of choosing a good
binary labeling for \gls{apsk} constellations in the presence of
\gls{nlpn}. We pay special attention to a class of \gls{apsk}
constellation which allows the use of a Gray labeling, which we call
rectangular \gls{apsk}. For this class, we propose a method to choose
the phase offsets of the constellation rings resulting in near-optimal
performance. The proposed method might also be useful when soft
information is passed to a decoder in the form of bit-wise
log-likelihood ratios in a \gls{bicm} scheme. For \gls{bicm}, it is
known that the labeling can have a significant impact on the
achievable information rate and the system performance
\cite{Caire1998}.



The remainder of the paper is organized as follows. In \MYsectionname
\ref{sec:system_model}, we present the channel model and define the
generic \gls{apsk} signal set. In \MYsectionname \ref{sec:detection},
we briefly review \gls{ml} detection and then describe and analyze the
suboptimal \gls{ts} detector together with the corresponding
\gls{pdf}. The results of the constellation optimization with respect
to \gls{sep} are presented and discussed in \MYsectionname
\ref{sec:optimization}. Binary labelings are discussed in
\MYsectionname \ref{sec:labelings}. Concluding remarks can be found in
\MYsectionname \ref{sec:conclusion}. 

\section{System Model}
\label{sec:system_model}



\subsection{Channel}

We consider the discrete memoryless channel \cite[Ch.  5]{Ho2005}
\begin{equation}
	Y = (X + Z) e^{-\imag \NLPN},
	\label{eq:channel}
\end{equation}
where $\imag \define \sqrt{-1}$ denotes the imaginary unit, $X \in
\setX$ the complex channel input, $\setX$ the signal constellation,
$Z$ the total additive noise, $Y$ the channel observation, and $\NLPN$
the \gls{nlpn}, which is given by \cite[Ch.  5]{Ho2005}
\begin{equation}
	\NLPN = \frac{\gamma \Len}{\NumSpans} \sum_{i=1}^{\NumSpans} |X +
	Z_i|^2. 
	\label{eq:nlpn}
\end{equation}
In \eqref{eq:nlpn}, $\gamma$ is the nonlinear Kerr-parameter, $\Len$ is the total
length of the fiber, $\NumSpans$ denotes the number of fiber segments,
and $Z_{i}$ is the noise contribution of all fiber segments up to
segment $i$. More precisely, $Z_i \define N_1 + \ldots + N_{i}$ is
defined as the sum of $i$ independent and identically distributed
complex Gaussian random variables with zero mean and variance
$\sigma_0^2$ per dimension (real and imaginary parts). The
total additive noise of all $\NumSpans$ fiber segments is denoted by $Z \define
Z_{\NumSpans}$, which has variance $\sigma^2 \define \E{|Z|^2} = 2
\NumSpans \sigma_0^2$, where $\E{\cdot}$ is the expected value. For
ideal distributed amplification, we consider the case $\NumSpans \to
\infty$. The total noise variance can be calculated as $\sigma^2 = 2
n_{sp} h \nu \alpha \Delta \nu  \Len$ \cite[Sec.
IX-B]{Essiambre2010}, where all parameters are taken from
\cite{Lau2007d} and are summarized in Table \ref{tab:constants}.
The additive noise power spectral density as defined in
\cite{Essiambre2010} is then given by $N_0 \define n_{sp} h \nu
\alpha =  \unit{1.04\cdot 10^{-20}}{\watt\per\kilo\metre\per\hertz}$.
Note that the total additive noise variance scales linearly with the
fiber length. 
\begin{table}[t]
	\caption{Constants and Parameters taken from \cite{Lau2007d}}
	\centering
	\begin{tabular}{ c | c | l }
		\bfseries symbol& \bfseries value & \bfseries meaning \\\hline \hline
		$\gamma$			 	& \unit{1.2}{\reciprocal\watt\kilo\reciprocal\metre} & nonlinearity parameter \\
		$n_{sp}$ 				& 1.41					 	& spontaneous emission factor \\
		$h$ 						& \unit{6.626\cdot 10^{-34}}{\joule\usk\second} & Planck's constant\\
		$\nu$ 					& \unit{1.936\cdot 10^{14}}{\hertz} & optical carrier frequency\\
		$\alpha$			 	& \unit{0.0578}{\kilo\reciprocal\metre} & fiber loss  (\unit{0.25}{\deci\bel\per\kilo\metre}) \\
		$\Delta \nu$	 	& \unit{42.7}{\giga\hertz} & optical bandwidth \\ \hline \hline
	\end{tabular}
	\label{tab:constants}
\end{table}

\begin{figure}[t]
	\begin{center}
		\includegraphics{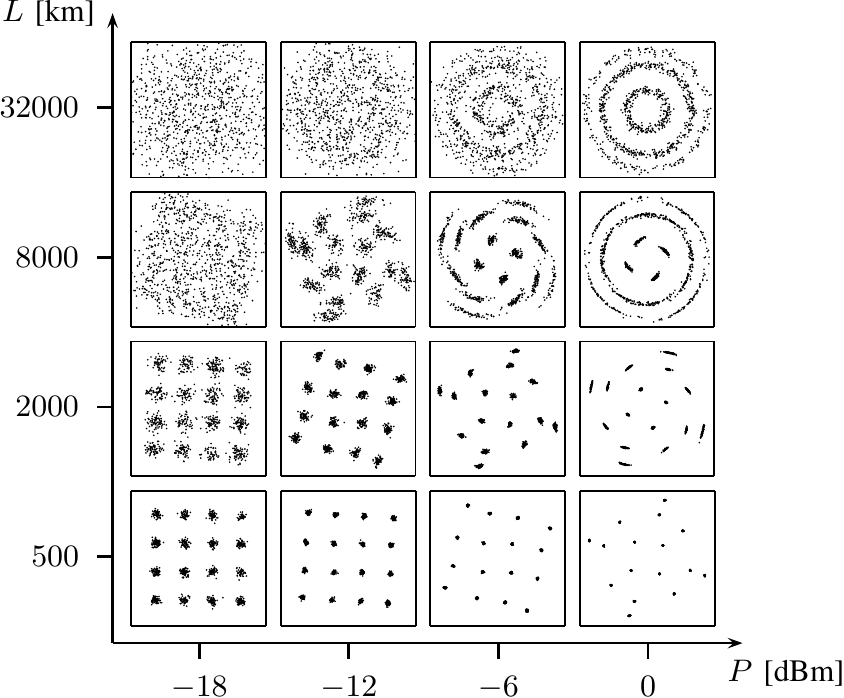}
	\end{center}
	\caption{Scatter plots of $Y$ assuming $X \in \setX_{\text{16-QAM}}$ for
	several combinations of transmit power $\Power$ and fiber length
	$\Len$.}
	\label{fig:signalnoise}
\end{figure}

An important aspect of this channel model is the fact that the
variance of the phase noise is dependent on the channel input (cf.
\eqref{eq:nlpn}), or equivalently on the average transmit power
$\Power$, defined as $\Power \define \E{|X|^2}$. In Fig.
\ref{fig:signalnoise} we show received scatter plots for $Y$ (cf.
\eqref{eq:channel}) assuming $X \in \setX_{\text{16-QAM}}$, where
\mbox{$\setX_{\text{16-QAM}} \define \{ \sqrt{P/10} (a+\imag b)
\forwhich a,b \in \{\pm 1, \pm 3 \}\}$} is the 16-QAM constellation,
and $\NumSpans=100$ for several combinations of $\Power$ and $\Len$.
The purpose of Fig. \ref{fig:signalnoise} is to gain insight into the
qualitative behavior of the channel. 
It can be observed that for very low input power and fiber length,
nonlinearities are negligible and the channel behaves as a standard
\gls{awgn} channel.  The scatter plots along a diagonal in Fig.
\ref{fig:signalnoise} correspond to a constant \gls{snr}, defined as
$\SNR \define {\Power}/{\sigma^2}$. In contrast to an \gls{awgn}
channel for which the scatter plots along any diagonal would look
similar, the received constellation points in Fig.
\ref{fig:signalnoise} start to rotate in a deterministic fashion and
the effect of the \gls{nlpn} becomes pronounced for large $\Len$ and
$\Power$. Therefore, in order to specify the operating point of the
channel, the \gls{snr} alone is not sufficient, because the parameter
space of the channel is two-dimensional, cf. \cite[Sec.
VII]{Yousefi2011a}.  In this paper, we present performance
results assuming a fixed fiber length and variable transmit power.

%
%
%
%

\subsection{Amplitude-Phase Shift Keying}
\label{sec:apsk}

We use the term \gls{apsk} for the discrete-input constellations
considered in this paper and focus on constellations with $M=4$, $8$,
and $16$ points. The \gls{apsk} signal set is defined as 
\cite{DeGaudenzi2006a}
\begin{equation}
	\setX \define \left\{
	r_k e^{\imag \left(\frac{2\pi j}{\ppr_k} + \phaseoffset_k \right)}
	\forwhich 1 \leq k \leq \NumRings , \quad 0 \leq j \leq \ppr_k - 1  \right\},
	\label{eq:apsk}
\end{equation}
where $\NumRings$ denotes the number of amplitude levels or rings,
$r_k$ the radius of the $k$th ring, $\ppr_k \geq 1$ the number of
points in the $k$th ring, where $\sum_{k=1}^{\NumRings} \ppr_k = M$,
and $\phaseoffset_k$ the phase offset in the $k$th ring.  Throughout
the paper, we assume a uniform distribution on the channel input $X$
over all symbols, and thus $\Power = (1/M)\sum_{k=1}^{\NumRings}
\ppr_k r_k^2$. The radii are assumed to be ordered such that $r_1 <
\ldots < r_{\NumRings}$ and we use $\vect{r} \define (r_1, \ldots,
r_{\NumRings})$ to denote the \emph{radius vector}. In this paper, for
$\ppr_1 = 1$, the point in the first ring is always placed at the
origin, implying $r_1 = 0$.  The radius vector is said to be uniform
if $r_{k+1} - r_k = \Delta$ for $1 \leq k \leq \NumRings-1$, where
$\Delta = r_2$ if $\ppr_1 = 1$ and $\Delta = r_1$ if $\ppr_1 \geq 2$.
The symbols are assumed to be indexed, \IE $x_i \in \setX$, $i =
1,\ldots,M$. The indexing is done starting in the innermost ring
($k=1$) by increasing $j$ from $0$ to $\ppr_1-1$ and then moving to
the next ring increasing $j$ from $0$ to $\ppr_2-1$ and so on. Thus,
finally we have $x_1 = r_1 e^{\imag \phaseoffset_1}, \ldots,x_M =
r_{\NumRings} e^{\imag 2 \pi (\ppr_{\NumRings-1})/\ppr_{\NumRings} +
\phaseoffset_{\NumRings}}$.

We also define the vectors $\vect{l} \define (\ppr_1, \ldots,
\ppr_{\NumRings})$ and $\vect{\phaseoffset} \define (\phaseoffset_1,
\ldots, \phaseoffset_{\NumRings})$, and use the notation
$\vect{l}$-APSK for an \gls{apsk} constellation with $\NumRings$ rings
and $\ppr_k$ points in the $k$th ring, \EG (4,4,4,4)-APSK. Note that
this notation does not specify a particular constellation without
ambiguity, due to the missing information about the radii and phase
offsets. 

\section{Detection Methods}
\label{sec:detection}

\subsection{Symbol Error Probability}

Let $\Reg_i$, $1\leq i \leq M$, be the decision region implemented by
a detector for the symbol $x_i$, \IE $\hat{X} = x_i$ if $Y \in
\Reg_i$, where $\hat{X}$ denotes the detected symbol. The average
\gls{sep} is then 
\begin{equation}
	\SEP = 1 - \frac{1}{M} \sum_{i = 1}^{M} P_{i \to i},
	\label{eq:sep}
\end{equation}
where $P_{i\to j} \define \text{Pr}[\hat{X} = x_j | X = x_i]$, $1 \leq
i, j \leq M$, are the symbol transition probabilities\footnote{For the
\gls{sep}, only the cases $j=i$, $1 \leq i \leq M$, need to be
considered. In Sec. \ref{sec:labelings}, all transition probabilities
are used. }
calculated as
\begin{equation}
	P_{i\to j} = \int_{\Reg_j} f_{Y|X=x_i} (y) \, \D y. 
	\label{eq:stp}
\end{equation}
That is, $P_{i\to j}$ is obtained through integration of the
conditional \gls{pdf} of the observation given the channel input $X =
x_i$ over the detector region for $x_j$.

\subsection{Maximum Likelihood Detection}


Let the polar representation of the channel input and the observation
be given by $X = \TXR e^{j\TXP}$ and $Y = \RXR e^{j\RXP}$,
respectively. The \gls{pdf} of $Y$ can be written in the form
\cite[Sec. III]{Yousefi2011a}, \cite[p.
225]{Ho2005}, \cite{Lau2007d}
\begin{equation}
	f_{Y|X=x} (y) = 
		\frac{f_{\RXR|\TXR= \txR}(\rxR)}{2 \pi \rxR} + \frac{1}{\pi \rxR}
	\sum_{m=1}^{\infty}
	\Re\left\{C_m(\rxR, \txR) e^{\imag m (\rxP - \txP)}\right\},
	\label{eq:pdf}
\end{equation}
where $x = \txR e^{\imag \txP}$, $y = \rxR e^{\imag \rxP}$, $\Re\{z\}$
is the real part of $z \in \complex$,  and
the \gls{pdf} of the received amplitude $\RXR$ given the transmitted
amplitude $\TXR=\txR$ is 
\begin{equation}
	f_{\RXR|\TXR=\txR}(r) = %
	\frac{2 \rxR}{\sigma^2} \exp\left(-\frac{\rxR^2+\txR^2}{\sigma^2}\right) %
	\bessel[0]{\frac{2 \rxR \txR}{\sigma^2}},%
	\label{eq:rice}
\end{equation}
where $\bessel[0]{\cdot}$ is the modified Bessel function of the first
kind. Analytical expressions for the coefficients $C_m(r, r_0)$ can be found in
\cite[Sec. III]{Yousefi2011a}. 
The \gls{ml} detector can now be described in the form of decision regions
$\Reg^{\text{ML}}_i \subset \complex$ for each symbol $x_i
\in \setX$:
\begin{equation}
 \Reg^{\text{ML}}_i\define 
 \bigcap_{\substack{j = 1\\ j \neq i}}^{M} 
 \{ y \in \complex \forwhich f_{Y|X=x_i}(y) \geq
f_{Y|X=x_j}(y) \}.
\label{eq:reg_ml}
\end{equation}
If we take the \gls{ml} decision regions defined in \eqref{eq:reg_ml},
then \eqref{eq:sep} becomes a lower bound on the achievable \gls{sep}
with suboptimal detectors. 

Evaluating the \gls{sep} by numerically integrating over the \gls{pdf}
\eqref{eq:pdf} is computationally expensive. Moreover, unlike for an
\gls{awgn} channel, where the \gls{ml} regions simply scale
proportionally to $\sqrt{\Power}$, the \gls{ml} decision regions
defined in \eqref{eq:reg_ml} change their shape based on the transmit
power $\Power$ \cite{Lau2007d}. This renders \gls{ml} detection
rather impractical for the purpose of constellation optimization.


\subsection{Two-Stage Detection}

In this paper, we study a slightly modified version of the suboptimal
\gls{ts} detector proposed in \cite{Lau2007d}. This detector is
a practical alternative to the \gls{ml} detector because it has much
lower complexity. In particular, the \gls{ts} detector employs
one-dimensional decisions: First in the amplitude direction (first
detection stage), followed by a phase rotation, and then in the phase
direction (second detection stage). 
\begin{figure}[t]
	\begin{center}
		\includegraphics{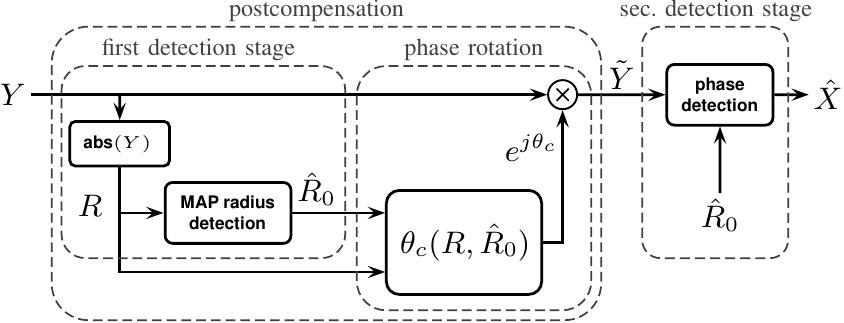}
	\end{center}
	\caption{Block diagram of the \gls{ts} detector. Note that the depicted
	postcompensation of $Y$ to $\tilde{Y}$ is reversible. }
	\label{fig:postcompensation}
\end{figure}

In Fig. \ref{fig:postcompensation}, we show a block diagram of the
\gls{ts} detector. We refer to the first detection stage together with
the phase rotation as \emph{postcompensation}. Based on the absolute
value of the observation $R$, radius detection is performed.  In
contrast to \cite{Lau2007d} and \cite{Beygi2011}, we use
\gls{map} instead of \gls{ml} radius detection and make use of the a
priori probability for selecting a certain ring at the transmitter,
thereby achieving a small performance advantage.  The radius detector
implements the rule: Choose $\TXRdet = r_k$, when $\mu_{k-1} \leq \RXR
< \mu_{k}$, where $\mu_k$, $0 \leq k \leq \NumRings$, denote the
decision radii or thresholds.  The \gls{map} decision threshold $\mu_k$, $1 \leq
k \leq \NumRings-1$, between $\ri$ and $\rii$ is obtained by solving
\begin{equation}
	\Pr{\TXR = \ri} f_{\RXR|\TXR=\ri}(\mu_k) = \Pr{\TXR = \rii}
	f_{\RXR|\TXR=\rii}(\mu_k),
	\label{eq:map_threshold}
\end{equation}
where the a priori probabilities are given by $\Pr{\TXR = \ri} =
\ppr_k/M$.\footnote{Solving \eqref{eq:map_threshold} for $\mu_k$ can
be done numerically and for an approximate analytical solution
assuming that $r_k \neq 0$, one may apply the high-SNR approximation
$\bessel[0]{x} \approx \frac{e^x}{\sqrt{2 \pi x}}$ as was done in
\cite{Beygi2011} for the \gls{ml} radius detector.} We always define
$\mu_0 \define 0$ and $\mu_{\NumRings} \define \infty$.
Based on the radius $\TXRdet$ of the detected ring and the
received amplitude $\RXR$, a correction angle $\theta_c$ is
calculated, by which the observation $Y$ is rotated to obtain the
postcompensated observation $\tilde{Y}$ as shown in Fig.
\ref{fig:postcompensation}.  The correction angle is given by
\begin{equation}
	\theta_c(\RXR, \TXRdet) = - \angle C_1(\RXR, \TXRdet),
	\label{eq:theta_cc}
\end{equation}
which is approximately a quadratic function in $\RXR$
\cite{Lau2007d}. 

The second detection stage is then performed with respect to
$\tilde{Y}$: A phase detector chooses the constellation point with
radius $\TXRdet$ that is closest to $\tilde{Y}$. Graphically, the
\gls{ts} detector employs so called annular sector regions (or annular
wedges) as decision regions for $\tilde{Y}$. 

\subsection{PDF of the Postcompensated Observation}
\label{subsec:pdf}

It is shown in \cite{Lau2007d} that for \gls{psk} signal sets
(which, in this paper, are denoted by $(M)$-APSK) where $\TXRdet
=\sqrt{\Power} = \text{const.}$, a minimum-distance detector for
$\tilde{Y}$ is equivalent to \gls{ml} detection. In contrast, for
constellations with multiple amplitude levels, the receiver structure
in Fig.  \ref{fig:postcompensation} does not perform \gls{ml}
detection.  However, in principle, {optimal} detection of $X$ is
still possible based on $\tilde{Y}$ due to the fact that the
postcompensation in Fig. \ref{fig:postcompensation} is invertible and
every invertible function forms a sufficient statistic for detecting
$X$ based on $Y$ \cite[p.  443]{Lapidoth2009}. Thus, the
performance loss associated with the \gls{ts} detection scheme
originates solely from suboptimal detection regions, not from the
postcompensation itself, which is a lossless
operation.\footnote{An important question that we do not address is
whether the phase rotation \eqref{eq:theta_cc} is still the best
choice for multilevel constellations, assuming that one is constrained
to straight-line phase decision boundaries for $\tilde{Y}$.}

In the following, we show how the \gls{pdf} of the postcompensated
observation $\tilde{Y}$ can be computed. This \gls{pdf} can then be
used to find optimal detection regions for $\tilde{Y}$. It is clear
from the block diagram of Fig. \ref{fig:postcompensation} that the
\gls{pdf} can be written as
\begin{equation}
	f_{\tilde{Y}|X=x}(\tilde{y}) = f_{Y|X=x}\left(\tilde{y} \cdot e^{-\imag
	\theta_c(\RXR,\TXRdet
	)}\right).
	\label{eq:pdf_pc}
\end{equation}
Most importantly, the correction angle $\theta_c$ is a discontinuous
function of the amplitude $\RXR$ because it depends on the detected
ring $\TXRdet$. In general, the correction angle can be written as
\begin{equation}
	\theta_c(\RXR, \TXRdet) = 
	\begin{cases}
		\theta_c(\RXR, r_{1}) \quad \text{if } \mu_{0} \leq \RXR < \mu_{1} \\
		\qquad \vdots \qquad \qquad \qquad  \vdots \\
		\theta_c(\RXR, r_{\NumRings}) \quad \text{if }\mu_{\NumRings-1} \leq \RXR < \mu_{\NumRings}\\
	\end{cases}.
	\label{eq:theta_c}
\end{equation}

%
%
%

\begin{figure}[t]
	\begin{center}
		\includegraphics{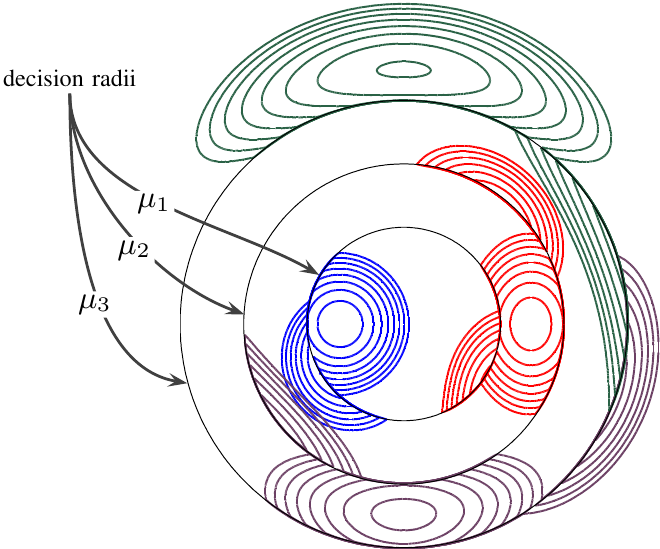}
	\end{center}
	\caption{PDF of $\tilde{Y}$ for $\Power = -\dBm{5}$ and $\Len =
	\unit{5500}{\kilo\metre}$ conditioned on 
	one particular point in each ring of the uniform (4,4,4,4)-APSK
	constellation. Color is helpful.}
	\label{fig:pdfsliced}
\end{figure}

For illustration purposes, we plot in Fig. \ref{fig:pdfsliced} the
\gls{pdf} resulting from \eqref{eq:pdf_pc} and \eqref{eq:theta_c}
conditioned on one particular point in each ring of the uniform
$(4,4,4,4)$-APSK constellation\footnote{The PDFs of the points which
are not shown look identical to the \gls{pdf} of the corresponding
point in the same ring up to a phase rotation.}. If we consider the
\gls{pdf} conditioned on $X = r_2$, \IE $\TXR = r_2$ and $\TXP = 0$
(shown in red), it can be observed that the contour lines look as
though they have been sliced up along the decision radii of the radius
detector. For $\RXR < \mu_1$, the correction angle is calculated with
respect to $r_1$, and thus, the phase is undercompensated. On the
other hand, for $\RXR \geq \mu_2$, the wrongly detected radius
results in an overcompensation. 

\subsection{Performance Comparison}

\begin{figure*}[t]
	\begin{center}
		\subfloat[ML decision regions w.r.t.\ $Y$]{
		\includegraphics[angle=180, origin=c, width=5.5cm]{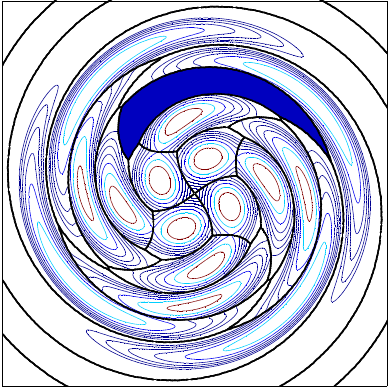}
		\label{fig:regions1}
		}
		\quad
		\subfloat[ML decision regions w.r.t.\ \smash{$\tilde{Y}$}]{
		\includegraphics[width=5.5cm]{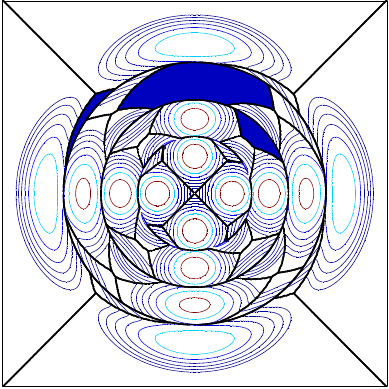}
		\label{fig:regions2}
		}
		\quad
		\subfloat[TS decision regions]{
		\includegraphics[width=5.5cm]{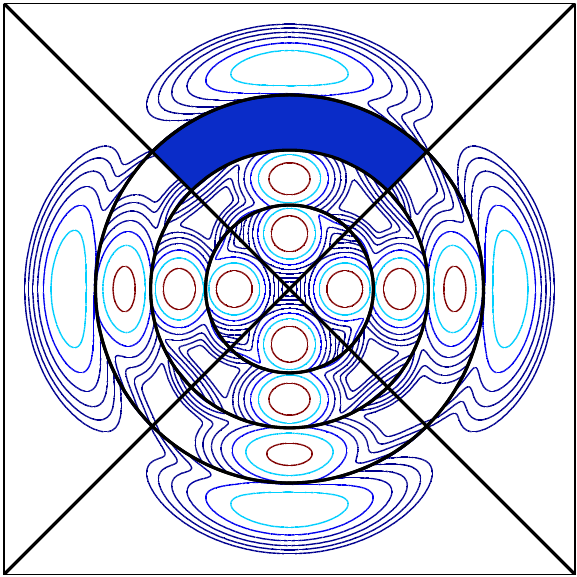}
		\label{fig:regions3}
		}
	\end{center}
	\caption{Decision regions for the uniform $(4,4,4,4)$-APSK
	constellation at $\Power = \dBm{-4}$. Shaded regions 
	correspond to the point $X = r_3 e^{\imag \pi/2}$. }
	\label{fig:regions}
\end{figure*}

For a qualitative performance comparison between the different
detectors, in Fig.~\ref{fig:regions}\subref{fig:regions1}, the
\gls{pdf} in \eqref{eq:pdf} is plotted for the uniform
$(4,4,4,4)$-APSK constellation together with the \gls{ml} decision
regions. In Fig.  \ref{fig:regions}\subref{fig:regions2}, the
\gls{pdf} in \eqref{eq:pdf_pc} is used instead. Finally, Fig.
\ref{fig:regions}\subref{fig:regions3} shows the same \gls{pdf} as
Fig. \ref{fig:regions}\subref{fig:regions2} together with the
suboptimal decision regions implemented by the \gls{ts} detector.  As
an example, the decision regions corresponding to the point $X = r_3
e^{\imag \pi/2}$ are shaded in Fig. \ref{fig:regions}. 

Comparing the
optimal decision regions in Fig.
\ref{fig:regions}\subref{fig:regions2} with the decision regions in
Fig.  \ref{fig:regions}\subref{fig:regions3}, it can be seen that
\gls{ts} detection is clearly suboptimal for this constellation and
input power. However, one would expect the two small shaded regions in
Fig. \ref{fig:regions}\subref{fig:regions2} to become smaller for
higher power. Intuitively, this is explained by the increasing
accuracy of the radius detector for increasing transmit power, due to
the Rice distribution \eqref{eq:rice} of the amplitude. More
precisely, let $P^{(e)}_{k} \define \text{Pr}[\hat{R}_0 \neq R_0 | R_0
= r_k]$ be the probability of an error in the first detection stage,
given that a symbol in the $k$th ring is transmitted. Then
\begin{equation}
	P^{(e)}_{k} = 1 - 
	\left(
	\MarcumQfunc{\normr{k},\normmu{k-1}}
	- \MarcumQfunc{\normr{k},\normmu{k}}
	\right),
	\label{eq:r1}
\end{equation}
where $\MarcumQfunc{\cdot,\cdot}$ is the Marcum Q-function, $\normr{k}
\define \sqrt{2} r_k / \sigma$, and $\normmu{k} \define \sqrt{2} \mu_k
/ \sigma$.  It follows that the \gls{sep} under \gls{ts} detection
converges to the \gls{sep} under \gls{ml} detection for increasing
input power and any \gls{apsk} constellation with uniform radius
vector, since then $P^{(e)}_{k}$, $1 \leq k \leq \NumRings$, tends to
zero as $\Power$ increases.

\section{Constellation Optimization}
\label{sec:optimization}

\subsection{Problem Statement}

In this section, we optimize the parameters of \gls{apsk}
constellations by minimizing the \gls{sep} under \gls{ts} detection.
Formally, the optimization problem can be stated as: Given $M$,
$\Power$, and $\Len$,
\begin{align}
	\underset{\vect{l},\vect{r}}{\text{minimize}} \qquad &
	\text{SEP under TS detection} \label{eq:objective}\\ 
	\text{subject to} \qquad & 1\leq \NumRings \leq M \label{eq:constraint1} \\
	& \ppr_1 + \ldots + \ppr_{\NumRings} = M\label{eq:constraint2} \\
	& \ppr_1 r^2_1 + \ldots + \ppr_{\NumRings} r^2_{\NumRings}= P M
	\label{eq:constraint3} \\
	& \ppr_k \geq 1, \text{for } 1 \leq k \leq \NumRings \label{eq:constraint4}.
\end{align} 
The objective function can be computed analytically using
\eqref{eq:sep} with the \gls{pdf} \eqref{eq:pdf_pc} integrated over
the \gls{ts} detector regions \cite[Eq. (4)]{Beygi2010}. Note that the
phase offset vector $\vect{\phaseoffset}$ does not appear in the
minimization problem.  This is due to the fact that the \gls{sep}
under \gls{ts} detection does not change assuming a phase offset in
any of the constellation rings: The \gls{pdf} of $\tilde{Y}$ is simply
rotated by the phase offset, but so is the detector region itself, and
hence the integrals in \eqref{eq:stp} are not affected. 

It is instructive to begin by discussing two special cases of the
general optimization problem above. The first case is obtained when
$\vect{r}$ is assumed to be uniform and an optimization is performed
only over the number of points in each ring
$\vect{l}$, cf. \eqref{eq:objective}$-$\eqref{eq:constraint4}. The
optimization problem then becomes an integer program which can be
solved in an exhaustive fashion for the constellation sizes considered
in this paper. The number of ways to distribute $M$ constellation
points to $i$ rings is given by ${M-1 \choose i-1}$. At most, there
are $M$ rings, which gives a total of $\sum_{i=1}^{M} {M-1 \choose
i-1} = 2^{M-1}$ possibilities to choose $\vect{l}$. There are $8$, $128$,
and $32768$ possibilities for $4$, $8$, and $16$ points, respectively.
It is clear that such a brute-force approach becomes unfeasible for larger
constellations.  However, based on the obtained results, it might be
possible to devise more sophisticated search methods for $M > 16$, \EG
by neglecting unrealistic combinations a priori. 

The second special case is given by the optimization of the radius
vector $\vect{r}$ for a certain constellation with fixed
parameter $\vect{l}$. For this case, the optimization problem becomes
a nonlinear program. Due to the power constraint, the dimensionality of
the search space is $\NumRings-1$ if $\ppr_1 > 1$ and $\NumRings-2$ if
$\ppr_1 = 1$, respectively. By inspecting the target function, one can
verify that this problem is nonconvex, \IE a local optimum does not
necessarily imply a global solution. We tested different nonlinear
solvers and obtained very good solutions with the Nelder--Mead simplex
method \cite{Nelder1965}. Even though the solution is not
guaranteed to be the globally optimal, we verified the global
optimality for several constellations and several combinations of
input power and fiber length with a brute-force grid search. 

By combining the solutions to the two special cases above, a solution
to the general optimization problem can then be found as follows: For
each \gls{apsk} constellation with a certain fixed parameter
$\vect{\ppr}$ one determines the optimal radius vector via
the simplex method for a given input power and fiber length, and then
optimizes over all possible $\vect{l}$.  We call this approach
\emph{joint optimization}. 

\subsection{Results and Discussion}

\subsubsection{4 Points}

\begin{figure}[t]
	\begin{center}
		\includegraphics{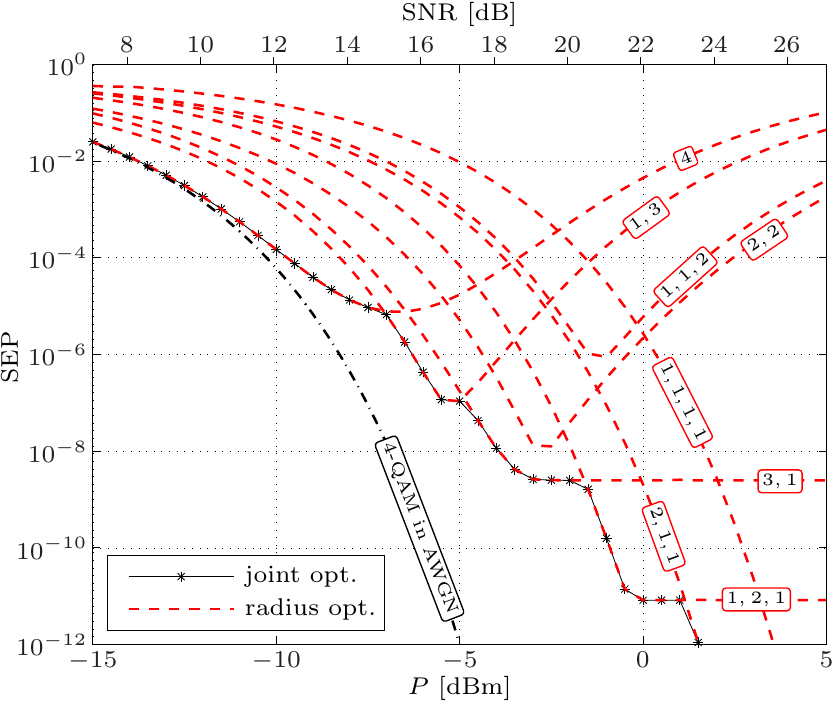}
	\end{center}
	\caption{Results for the constellation optimization with $M=4$. The
	fiber length is $L = \unit{7000}{\kilo\metre}$. The dash-dotted line is a
	reference curve, showing the \gls{sep} of $4$-QAM in
	an \gls{awgn} channel under \gls{ml} detection.}
	\label{fig:4points_joint_annotated}
\end{figure}

We start by finding optimal APSK constellations with $M=4$ points.
The fiber length is fixed at $L = \unit{7000}{\kilo\metre}$ and the
input power $\Power$ is varied from $\dBm{-15}$ to $\dBm{5}$ in steps
of $\dBm{0.5}$.  In Fig. \ref{fig:4points_joint_annotated} we plot the
performance of all possible eight 4-point APSK constellations with
optimal radius vector (dashed lines) and the curves are labeled with
the corresponding $\vect{\ppr}$. (For $(4)$-APSK and $(1,3)$-APSK the
radius vector is fixed, \IE no optimization is performed.) The results
of the joint optimization are shown with markers. We also show the
\gls{sep} of (4)-APSK (or 4-QAM) in an \gls{awgn} channel under
\gls{ml} detection as a well-known reference curve (dash-dotted line).
Note that for each $\Power$ the \gls{sep} is minimized by a
constellation with certain parameters $\vect{\ppr}$ and $\vect{r}$.
For example, it can be seen that for an input power range between
$\dBm{-15}$ and $\dBm{-7.5}$, $(4)$-APSK is optimal. 

Based on the behavior of the \gls{sep} for the individual \gls{apsk}
constellations with optimized radius vector (dashed lines in Fig.
\ref{fig:4points_joint_annotated}), it is possible to classify the
constellations into three classes. The first class exhibits a well
known performance minimum, \IE an optimal operating power.  The second
class does not exhibit a minimum, but eventually flattens for very
high input power (see, \EG the performance of $(3,1)$-APSK). The third
class exhibits a performance behavior which is strictly and steadily
decreasing with increasing input power.

\begin{figure}[t]
	\begin{center}
		\includegraphics{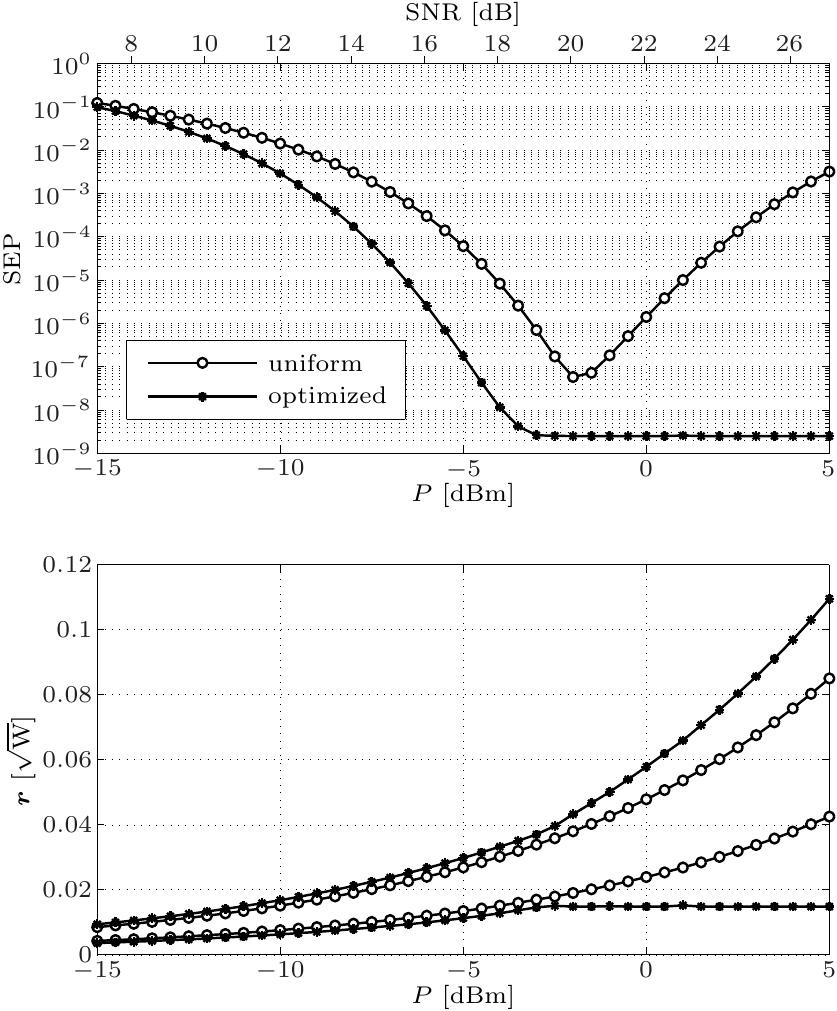}
	\end{center}
	\caption{Performance of the $(3,1)$-APSK constellation with a
	uniform and optimized radius vector (top) and the
	corresponding radius vector (bottom).}
	\label{fig:sep_flatten}
\end{figure}

The flattening of the \gls{sep} for the second class is explained by
the availability of a \emph{sacrificial point} in the outer
constellation ring. The meaning of the term \emph{sacrifical} is best
explained with the help of an example. In Fig.  \ref{fig:sep_flatten}
we show the results of the radius optimization for $(3,1)$-APSK (top),
together with the optimal values of the parameter $\vect{r}$ (bottom). 
It can be observed that, for $\Power > \dBm{-3}$, the optimal ring
spacing shows a very peculiar behavior. For this power regime, $r_1$
appears to be fixed and any increase in average power is absorbed by
putting the outermost point further away from the inner ring. In some
sense, the outer point (experiencing very high \gls{nlpn}) is
sacrificed with the result of saving the average \gls{sep} of the
constellation. Observe that $(1,1,1,1)$-APSK is the only \gls{apsk}
constellation with 4 points that belongs to the third
class\footnote{The \gls{sep} for $(2,1,1)$-APSK in Fig.
\ref{fig:4points_joint_annotated} flattens for very high input power.
} and it was already argued in \cite{Lau2007d}, that this
constellation is optimal for very high input power.\footnote{For
\gls{apsk} constellations with only one point in each ring the
\gls{sep} under \gls{ts} detection can be calculated using
\eqref{eq:r1} as $\SEP= \frac{1}{M} \sum_{k=1}^{M} P_k^{(e)}$.}



\begin{figure}[t]
	\begin{center}
		\includegraphics{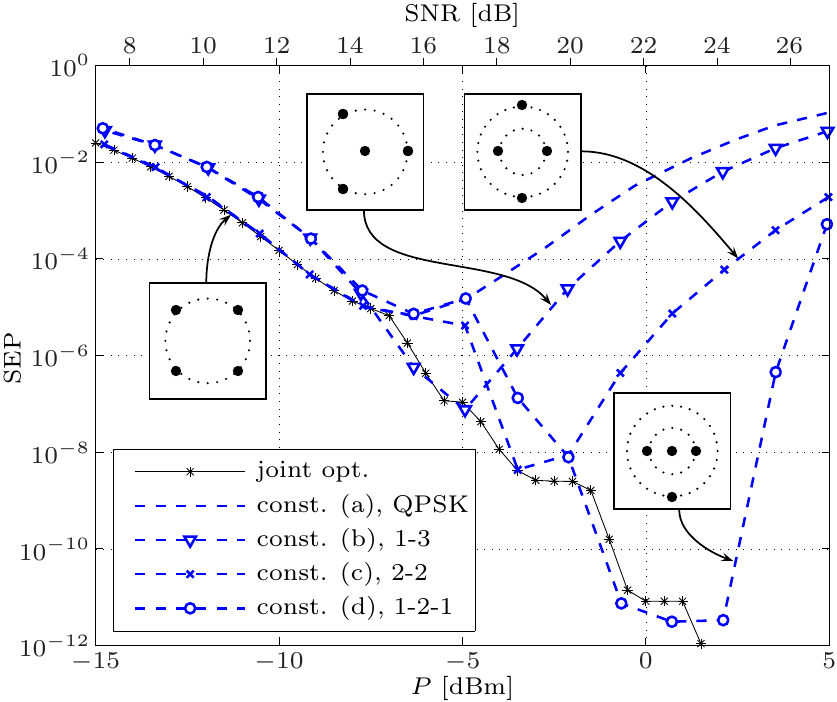}
	\end{center}
	\caption{Comparison between the joint optimization of the 4-point
	\gls{apsk} constellation and the results for optimized
	constellations (a)--(d) in \cite{Lau2007d} under \gls{ml}
	detection. The legend shows the constellation names originally used
	in \cite{Lau2007d}.   }
	\label{fig:LauKahn}
\end{figure}

The system parameters are chosen in such a way that the obtained
results are directly comparable to the performance of the optimized
constellations presented in \cite[Sec.  IV]{Lau2007d}. To
facilitate a comparison, in Fig. \ref{fig:LauKahn} we provide a
digitalized version of \cite[Fig. 15]{Lau2007d} and plot the
outcome of the joint optimization in the same figure. All four
constellations used in \cite{Lau2007d} can be seen as
\gls{apsk} constellations and they are depicted in Fig.
\ref{fig:LauKahn} for convenience. With the notation introduced in
this paper, the constellations are $(4)$-APSK, $(1,3)$-APSK,
$(2,2)$-APSK with $\vect{\phaseoffset} = (0, \phaseoffset_2)$, and
$(1,2,1)$-APSK with $\vect{\phaseoffset} = (0, 0, \phaseoffset_3)$.
The parameters $\phaseoffset_2$ and $\phaseoffset_3$ are determined by
a precompensation technique developed in \cite{Lau2007d}, while
the radius vector of the two latter constellations was
optimized. 
It is important to point out that the optimization in
\cite{Lau2007d} was performed with respect to \gls{ml}
detection, while for the joint optimization in this paper the
suboptimal \gls{ts} detector is assumed. Notice that, for $(4)$-APSK
these two detection schemes are equivalent and hence, the performance
results taken from \cite{Lau2007d} for $(4)$-APSK
(constellation $(a)$) in Fig. \ref{fig:LauKahn} overlap with the
results of the joint optimization for an input power up to \dBm{-7.5}.  
Comparing the results in Fig. \ref{fig:LauKahn}, it can be seen that
the jointly optimized \gls{apsk} constellations perform very close to
the optimized constellations in \cite{Lau2007d}. For certain
input powers, \EG \dBm{-7} or \dBm{-5}, a performance loss is visible,
which is explained by the weaker detection technique. On the other
hand, for some power regimes, \EG \dBm{-3.5} or \dBm{-2}, the jointly
optimized constellations perform as well as, or even outperform, the
best constellations presented in \cite{Lau2007d}. We attribute
this performance gain to the more systematic search which is offered
by the \gls{apsk} framework. Also note that there is no need to find
phase precompensation angles as was done in \cite{Lau2007d},
because those are relevant only for \gls{ml} detection, but irrelevant
for the performance under \gls{ts} detection. This removes one degree
of freedom from choosing a constellation and makes the optimization
simpler. As a last point, it is unclear why constellation (d) in
\cite{Lau2007d} does not exhibit a flattening \gls{sep} for
very high input power, even though the radius vector was optimized and
a sacrificial point is available. We conjecture that the \gls{sep}
results for constellation (d) for \dBm{3} and \dBm{5} are only locally
optimal. 


\subsubsection{8 Points}

\begin{figure}[t]
	\begin{center}
		\includegraphics{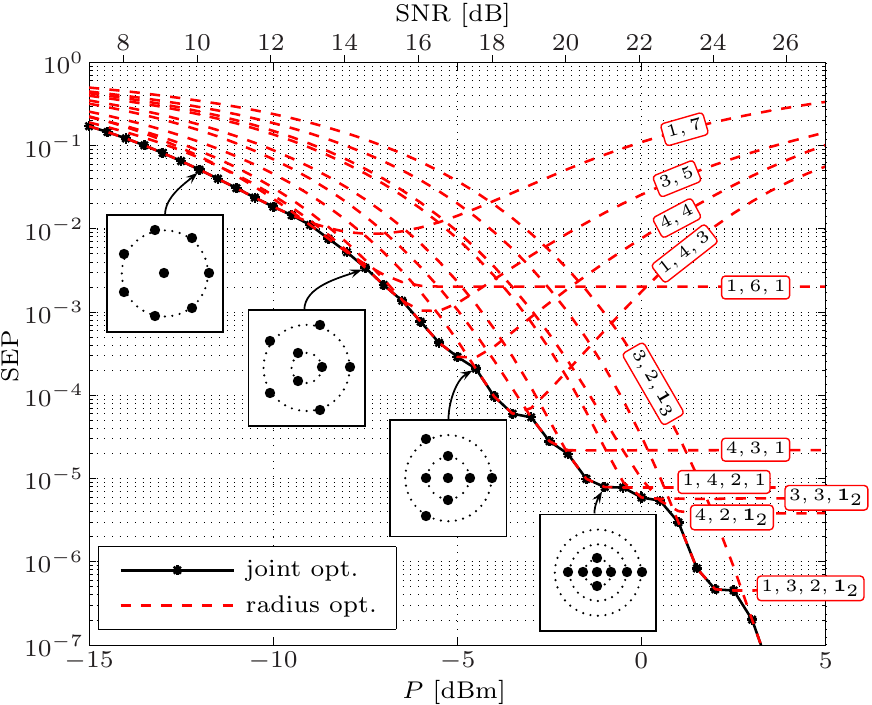}
	\end{center}
	\caption{
	Results for the constellation optimization with $M=8$. The
	fiber length is $L = \unit{7000}{\kilo\metre}$. }
	\label{fig:8points_joint_annotated}
\end{figure}

For $M=8$ points, we present optimization results for the same system
parameters as before. The results for the joint optimization are shown
in Fig. \ref{fig:8points_joint_annotated} with markers. 
Since it is not very instructive to show the performance of all
8-point \gls{apsk} constellations in the same figure, we only plot the
\gls{sep} of those constellations that are optimal somewhere in the
considered power range (dashed lines). Thus, the parameter
$\vect{\ppr}$ is indicated by the labeling of the corresponding dashed
line. To avoid cumbersome notation, we also define $\vect{1}_N \define
1,\ldots,1$ ($N$ times). To get a more intuitive feeling for the
optimal constellations in different power regimes, the inset figures
show the actual constellations that are optimal at \dBm{-12},
\dBm{-7.5}, \dBm{-4.5}, and \dBm{-1}. The constellations are shown
with their optimized radius vector for the corresponding input power.

\begin{figure}[t]
	\begin{center}
		\includegraphics{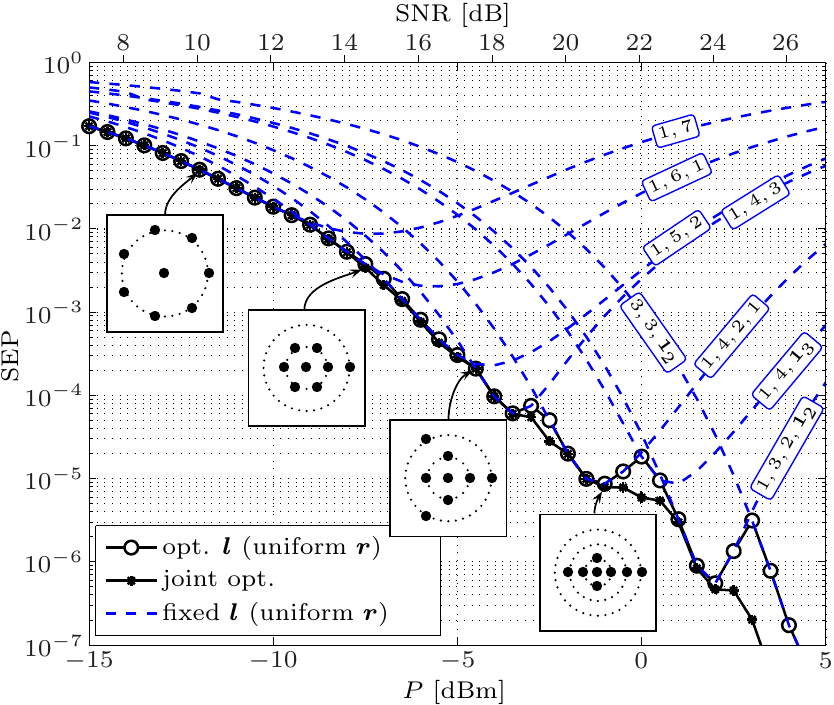}
	\end{center}
	\caption{
	Results for the constellation optimization with $M=8$ (white circles)
	assuming that a uniform radius vector for all constellations. The fiber length is $L =
	\unit{7000}{\kilo\metre}$. The results of the joint optimization
	from Fig. \ref{fig:8points_joint_annotated}
	(black markers) are
	shown for comparison. The dashed lines
	correspond to the \gls{sep} of the constellations that are
	indicated by the labels. }
	\label{fig:8points_ppr_annotated}
\end{figure}

We also perform an optimization only over $\vect{\ppr}$ assuming that
the radius vector is uniform. The results are depicted in Fig.
\ref{fig:8points_ppr_annotated}. To facilitate a better comparison
with the jointly optimized constellations, the solid black line in
Fig. \ref{fig:8points_joint_annotated} is again plotted in Fig.
\ref{fig:8points_ppr_annotated}. The dashed lines in Fig.
\ref{fig:8points_ppr_annotated} show the \gls{sep} of the \gls{apsk}
constellation with a fixed parameter $\vect{\ppr}$ and assuming a
uniform radius vector. An important observation here is that up to an
input power of \dBm{-3.5}, there is almost no difference between the
performance of the jointly optimized constellations and the optimal
constellations obtained assuming a uniform radius vector, suggesting
that most of the performance improvement is due to optimizing the
parameter $\vect{l}$. 

\subsubsection{16 Points}

\begin{figure}[t]
	\begin{center}
		\includegraphics{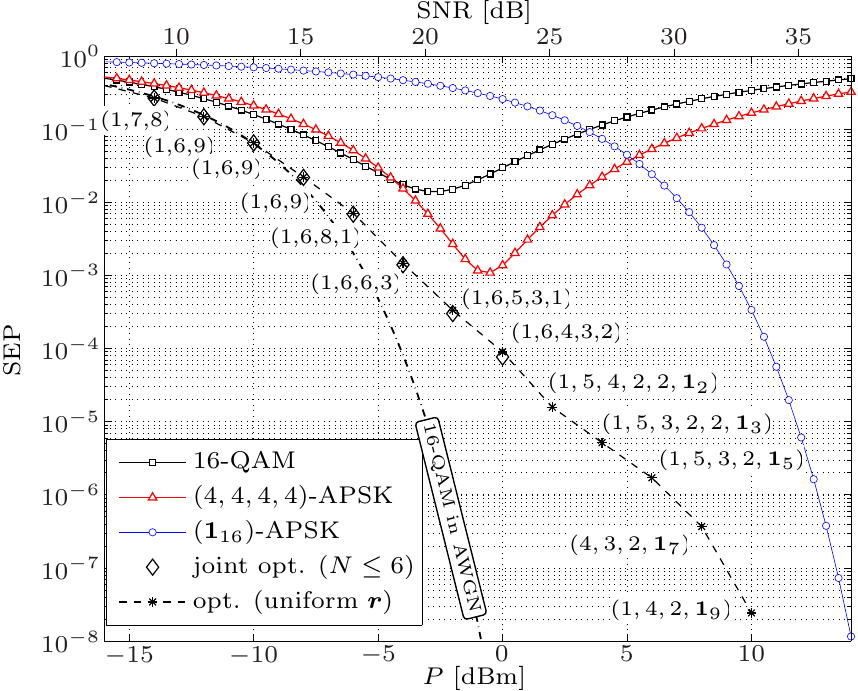}
	\end{center}
	\caption{Results for the constellation optimization with $M=16$. 
	The
	\gls{sep} under \gls{ts} detection for 16-QAM, uniform $(4,4,4,4)$-APSK, and
	$(\vect{1}_{16})$-APSK are shown for comparison. 
	The \gls{sep} under
	\gls{ml} detection for 16-QAM in an \gls{awgn} channel (dash-dotted
	line) is shown as a reference. }
	\label{fig:ppr_optimization}
\end{figure}

Motivated by the results obtained for $M=8$, for $M=16$ points, we
limit ourselves to the case where the radius vector is assumed to be
uniform for all constellations. The fiber length is $\Len =
\unit{5500}{\kilo\metre}$ and the input power $\Power$ is varied from
\dBm{-14} to \dBm{10} in steps of \dBm{2}. The results are shown in
Fig. \ref{fig:ppr_optimization}, where we indicate the optimal
parameter $\vect{l}$ next to the corresponding marker of the curve. 
The individual \gls{sep} under \gls{ts} detection for 16-QAM,
uniform $(4,4,4,4)$-APSK, and $(\vect{1}_{16})$-APSK are shown for comparison,
while the \gls{sep} under \gls{ml} detection for 16-QAM in an
\gls{awgn} channel is shown as a reference. The results in Fig.
\ref{fig:ppr_optimization} show that up to an input power of \dBm{-8},
the performance of the optimized constellations follows closely the
performance of 16-QAM in \gls{awgn}. In other words, for this power
regime it is possible to find \gls{apsk} constellations with
\gls{ts} detection that perform as well as 16-QAM for a channel
without nonlinear impairments under \gls{ml} detection. For higher
input power, the optimized constellations gradually utilize more
amplitude levels, due to the increase in \gls{nlpn}. If we take as a
baseline the minimum \gls{sep} achieved by the 16-QAM constellation
($\Power \approx \dBm{-2.8}$ and \gls{sep} $\approx 10^{-2}$), and
interpolate the optimal \gls{apsk} performance for the same \gls{sep},
we observe that a performance gain of \dBm{3.2} is achieved. 

In order to verify the assumption that a joint optimization approach
does not lead to significant performance gains, we also perform a
reduced complexity approach to the joint optimization, where the
optimization is restricted to constellations with at most $\NumRings =
6$ rings. This makes the results meaningful only for low and moderate
input power ($P \leq \dBm{0}$), because, as we have seen previously,
for higher input power, the dominance of the \gls{nlpn} dictates the
use of more amplitude levels to achieve good performance.  The results
are also shown in Fig. \ref{fig:ppr_optimization} for $\Power \leq
\dBm{0}$ (diamond markers). It can be seen that the jointly optimized
constellations with the additional constraint $\NumRings \leq 6$
follow closely the performance of the constellations obtained for a
uniform radius vector, confirming that the joint optimization only
yields negligible performance improvements for this power regime.  For
higher input power, the obtained jointly optimized constellations
perform worse than the optimal constellations obtained with a uniform
radius vector, which is simply due to the restriction to six rings.

\begin{figure}[t]
	\begin{center}
		\includegraphics{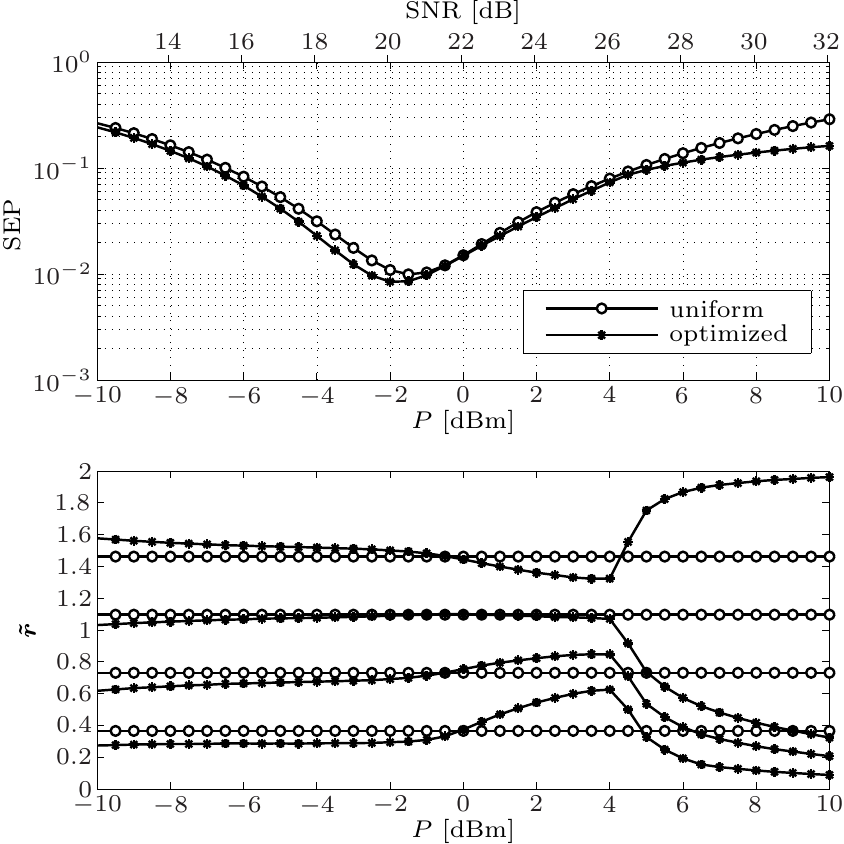}
	\end{center}
	\caption{Illustration of a sacrificial ring which occurs for the
	radius optimization of $(4,4,4,4)$-APSK for $\Power > \dBm{4}$. The
	system length is $\Len = \km{7000}$.}
	\label{fig:radii_opt4444}
\end{figure}

Finally, the phenomenon of sacrificial points discussed previously
also generalizes to entire rings, \IE when optimizing the radius
vector of constellations with more than one point in the outer ring.
In this case, however, the \gls{sep} still exhibits a minimum.  As an
example, in Fig. \ref{fig:radii_opt4444} we show the result of the
radius optimization for $(4,4,4,4)$-APSK (top) together with the
optimal radius vector (bottom). The radius vector $\vect{r}$ is
plotted in a normalized fashion $\vect{\tilde{r}} =
\vect{r}/\sqrt{\Power}$ (\IE $\sum_{i=1}^{\NumRings} \ppr_i
\tilde{r}_i^2 = 1$), so that the effect is more clear. It can be
observed that up to an input power of \dBm{4} the distance between any
two adjacent rings for the optimal radius vector is approximately the
same.  Moreover the distance decreases with higher input power (like
``squeezing accordion pleats'' \cite{Freckmann2009}).  However,
for $P > \dBm{4}$ the optimal radius vector changes significantly.
Somewhat counterintuitively, in this power regime, is is better to
place the four points in the outer constellation ring far away from
the other rings. 



\section{Binary Labelings}
\label{sec:labelings}

In order to allow for the transmission of binary data, we assume that
each symbol $x_i \in \setX$ is labeled with a binary vector
$\vect{c}_i = (c_{i,1}, \ldots, c_{i,m}) \in \{0,1\}^m$, where $m =
\log_2 M$. The different binary vectors are the binary representations
of the integers $\{0,1,\dots, M-1\}$.\footnote{One may arbitrarily
choose $c_{i,1}$ as the most significant bit.} A specific mapping
between vectors and constellation symbols is called a binary labeling,
which will be denoted by an $M \times m$ matrix $\labeling_m =
(\vect{c}_1^T, \ldots, \vect{c}_M^T)^T$. 

A \emph{Gray labeling} is obtained if the binary vectors of
neighboring symbols, \IE symbols that are closest in terms of
Euclidean distance, differ by only one bit position. As an example,
the Gray labeling $\BRGC_m$ for $(M)$-\gls{apsk}
constellations\footnote{Different Gray labelings exist for a given
constellation and for simplicity we restrict ourselves to $\BRGC_m$,
which is referred to as the \gls{brgc} in the literature: It is the
provably optimal Gray labeling for \gls{psk} constellations in an
\gls{awgn} channel at high \gls{snr} \cite{Agrell2004a}. } may be
constructed by $m-1$ recursive reflections of the trivial labeling
$\BRGC_1 = (0,1)^T$. To obtain $\BRGC_{m+1}$ from $\BRGC_{m}$ by
reflection, generate the matrix $(\vect{c}_1^T, \ldots, \vect{c}_M^T,
\vect{c}_M^T, \ldots, \vect{c}_1^T)^T$ and add an extra column from
the left, consisting of $M$ zeros followed by $M$ ones
\cite{Agrell2004a}.


\subsection{Bit Error Probability}
\label{subsec:bep}

The average \gls{bep} of the signal constellation is given by
\begin{align}
	\text{BEP} = \frac{1}{m M} \sum_{i=1}^{M} \sum_{j=1\atop j\neq i}^{M}
	d_{\text{H}}(\vect{c}_i, \vect{c}_j) \cdot
	P_{i \to j}, 
	\label{eq:bep}
\end{align}
where $d_{\text{H}}(\cdot, \cdot)$ denotes the Hamming distance
between two binary vectors. 
A lower bound for the \gls{bep} is $\text{SEP}/m \leq
\text{BEP}$, which directly follows from $1 \leq
d_{\text{H}}(\vect{c}_i, \vect{c}_j)$, $i \neq j$.


The probabilities $P_{i \to j}$ are fixed for a given constellation
and $\Power$ and $\Len$ (cf. \eqref{eq:stp}), hence \eqref{eq:bep}
depends only on the labeling. However, it is important to realize that
the phase offset vector $\vect{\phaseoffset}$ has an effect on these
probabilities for $j \neq i$. Therefore, even though two APSK
constellations with the same $\vect{\ppr}$ and $\vect{r}$ but
different $\vect{\phaseoffset}$ have the same \gls{sep}, they may have
a different \gls{bep}.  In the following, we show how to exploit this
new degree of freedom for a class of APSK constellations that permit
the use of a Gray labeling.

\subsection{Rectangular APSK}

\begin{figure}[t]
	\begin{center}
		\subfloat[Illustration in polar coordinates]{
			\includegraphics[width=5.5cm]{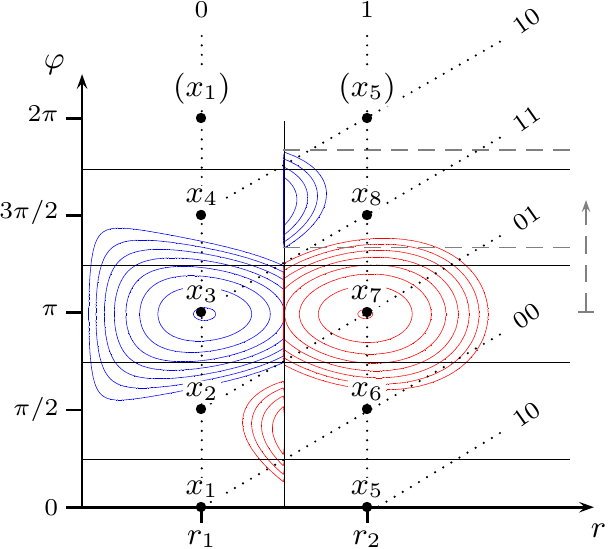}
		}
		\quad
		\subfloat[Labeled Const.]{
			\includegraphics{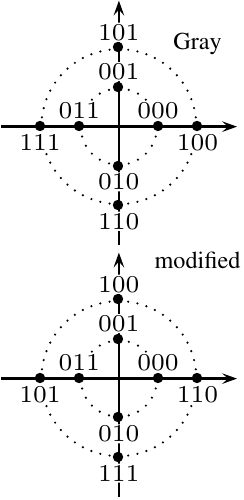}
		} 
	\end{center}
	\caption{In (a), the \gls{pdf} of $\tilde{Y}$ is shown for
	$(4,4)$-APSK conditioned
	on $X = x_3$ (blue) and $X = x_7$ (red) for $\Len =
	\unit{7000}{\kilo\metre}$ and $\Power = \dBm{-5}$. Solid lines
	correspond to the decision boundaries of the \gls{ts} detector and
	dotted lines show connections between neighboring symbols according
	to the \gls{pdf}. }
	\label{fig:labeling44annotated}
\end{figure}

APSK constellations with $1 < \NumRings < M$, $\vect{\phaseoffset} =
\boldsymbol{0}$ and $\ppr \define \ppr_k = M/\NumRings$, $1 \leq k
\leq \NumRings$, have a ``rectangular'' structure when plotted in
polar coordinates. For these constellations, a Gray labeling is given
by $\mathbb{L}_m = \BRGC_{\log_2 \NumRings} \otimes \, \BRGC_{\log_2
\ppr}$, where $\otimes$ is the ordered direct product, defined as
\begin{equation}
	(\vect{a}_1^T, \ldots, \vect{a}_p^T)^T \otimes (\vect{b}_1^T,
	\ldots, \vect{b}_q^T)^T = (\vect{c}_1^T,
	\ldots, \vect{c}_{pq}^T)^T, 
\end{equation}
where $\vect{c}_{q i + j} = (\vect{a}_i, \vect{b}_j)$, for $1 \leq i
\leq p$ and $1 \leq j \leq q$. This amounts to independently choosing
a Gray labeling for the radius and phase coordinates of the
constellation and then concatenating the binary vectors. In Fig.
\ref{fig:labeling44annotated}(b) (top) an example for such a
construction is shown. 


Gray labelings ensure that for a standard \gls{awgn} channel the
\gls{bep} closely approaches the lower \gls{bep} bound for high
\gls{snr} in Sec.\ \ref{subsec:bep}. Using the same labeling in a nonlinear channel, however,
does not necessarily ensure good performance since it is constructed
without considering the underlying \gls{pdf} of the observation. From
Fig.  \ref{fig:pdfsliced} it is evident that this \gls{pdf} has a
rather unusual shape, due to the slicing effect caused by the
postcompensation. To further illustrate this point, consider
$(4,4)$-APSK with $r_1/r_2 = 0.424$ and $\vect{\phaseoffset} = (0,0)$,
operating at $\Len = \unit{7000}{\kilo\metre}$ and $\Power =
\dBm{-5}$, which is the optimal \gls{apsk} constellation for these
parameters (cf. Fig.  \ref{fig:8points_joint_annotated}). The
\gls{pdf} of $\tilde{Y}$ conditioned on one point in each
ring\footnote{The \gls{pdf}s conditioned on all other points can be
obtained by a phase translation. Note that the PDF is periodic in
phase. } is evaluated and plotted \emph{in polar coordinates}, as
shown in Fig.  \ref{fig:labeling44annotated}(a) for $x_3$ and $x_7$. 
The solid lines correspond to the decision boundaries of the \gls{ts}
detector. Recalling that the symbol transition probabilities are
obtained through integration of the \gls{pdf} over the detector
regions (cf.  \eqref{eq:stp}), Fig. \ref{fig:labeling44annotated}(a)
can then be used to identify ``neighboring symbols'' of $x_3$ and
$x_7$ (and consequently of all points) in the sense that the
corresponding transition probabilities will dominate in
\eqref{eq:bep}. The main observation here is that, even though $x_3$
and $x_7$ are adjacent symbols in radial direction, the corresponding
transition probabilities, \IE $P_{3 \to 7}$ and $P_{7 \to 3}$, are
negligible compared to $P_{3 \to 8}$ and $P_{7 \to 2}$, respectively.
The dotted lines in Fig. \ref{fig:labeling44annotated}(a) show
appropriate connections between neighboring symbols taking into
account the nonlinear \gls{pdf}.
In Fig. \ref{fig:labeling44annotated}(b) the $(4,4)$-constellation is
shown with the Gray labeling $\BRGC_1 \otimes\, \BRGC_2$ (top) and the
modified labeling (bottom) that results from ``following'' the dotted
lines in Fig. \ref{fig:labeling44annotated}(a) and concatenating the
binary vectors.  

Observe that the labeled constellation in the bottom of Fig.
\ref{fig:labeling44annotated}(b) may be obtained from the Gray labeled
constellation by using a phase offset vector of $\vect{\phaseoffset} =
(0, \pi/2)$: In this case, the non-zero phase offset in the second
ring does not change the constellation (\IE the set of symbols), but
rather leads to a different indexing of symbols (cf. Sec.
\ref{sec:apsk}), and consequently to a different mapping between
symbols and binary vectors.  
Going one step further, we now allow for arbitrary phase offsets in
all constellation rings\footnote{Note that an \gls{apsk} constellation with
$\vect{\phaseoffset} \neq \vect{0}$ is not necessarily rectangular
anymore.}, with the intention to ``steer'' the phase decision
boundaries such that they are roughly symmetric around the \gls{pdf}.
Starting from a Gray labeled rectangular \gls{apsk} constellation, a
simple way to achieve this is by initializing $\phaseoffset_1 = 0$ and
then calculating
\begin{equation}
	\phaseoffset_i = \theta_c(\mu_{i-1}, r_{i-1}) - \theta_c(\mu_{i-1},
	r_{i}) + \phaseoffset_{i-1}, 
	\label{eq:phaseoffset}
\end{equation}
for $i = 2, \ldots, \NumRings$. For the previous example, evaluating
\eqref{eq:phaseoffset} for $i=2$ results in $\phaseoffset_2 \approx
1.878$, corresponding to the length of the dashed, grey arrow in
\ref{fig:labeling44annotated}(a). The two dashed, grey lines are the new
phase decision boundaries for $x_7$ and it can be seen that they
appear roughly symmetric around the blue \gls{pdf}. We highlight that
this proposed method to determine the phase offset vector may be
applied to any rectangular \gls{apsk} constellation of arbitrary
constellation size, provided that $M$ is a power of 2.

\subsection{Results and Discussion}

\begin{figure}[t]
	\begin{center}
		\includegraphics{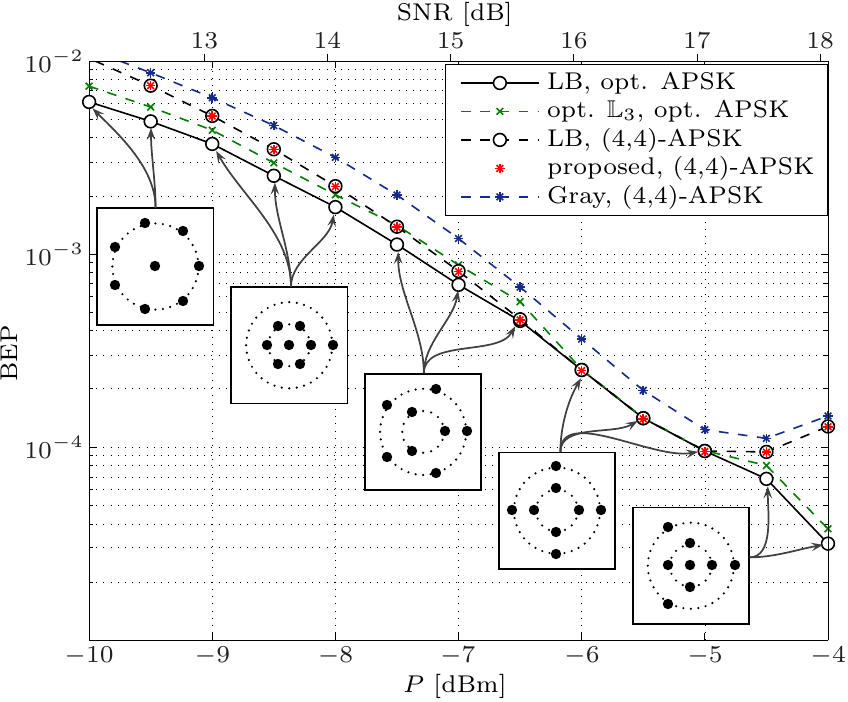}
	\end{center}
	\caption{Average \gls{bep} and lower bounds (LB) for $M = 8$ and
	$\Len = \km{7000}$ for different APSK constellations. The subfigures
	show the optimized \gls{apsk} constellations for the corresponding
	input power. }
	\label{fig:LabelingBEP}
\end{figure}

In Fig. \ref{fig:LabelingBEP}, the \gls{lb} for the \gls{bep} is
plotted for (4,4)-APSK with optimized $\vect{r}$ and $\Len =
\unit{7000}{\kilo\metre}$ (black, dashed line). The average \gls{bep}
of the labeled constellation is shown with the proposed phase offset
vector (red markers) and $\vect{\phaseoffset} = \vect{0}$, resulting in
the conventional Gray labeling (blue, dashed line). The performance
with the proposed method is almost indistinguishable from the lower
bound and a gain of approximately $\dB{0.4}$ is
visible at $\text{SEP} = 10^{-3}$ over the Gray labeling approach. 

Moreover, in Fig. \ref{fig:LabelingBEP} we plot the \gls{lb} based on
the jointly optimized \gls{apsk} constellations with $M = 8$ (cf. Fig.
\ref{fig:8points_joint_annotated}), where the subfigures are provided
to show the optimal parameter $\vect{l}$ for the different input
powers. For each input power, the optimal labeling is determined
exhaustively\footnote{It was pointed out in \cite{Huang2005} that, in
general, the labeling problem falls under the category of quadratic
assignment problems, and as such, is NP-hard.} and the corresponding
\gls{bep} is shown by the green line. Note that the \gls{lb} is tight
only for the rectangular $(4,4)$-APSK.
The results demonstrate that first optimizing the constellation with
respect to \gls{sep} and then choosing an optimal labeling does not
guarantee to give the best \gls{bep} performance. In particular, for
\dBm{-7} and \dBm{-6.5}, (4,4)-APSK with the proposed phase offset
vector achieves a lower \gls{bep} than the jointly optimized
constellations (with respect to \gls{sep}) with an optimal labeling.

As a last point, one might argue that the class of rectangular
\gls{apsk} constellations is not particularly interesting, because
they appear rarely as optimal \gls{apsk} constellations with respect
to \gls{sep} (\EG for $M=8$ they only appear in a small power range
and for $M=16$ they do not appear at all). However, the previous
results show that if we take the \gls{bep} as the main figure of
merit, rectangular \gls{apsk} constellations may be advantageous in
certain power regimes, because they can closely approach the lower
\gls{bep} bound. Moreover, as we described earlier, the proposed
labeling method is easily applicable to any constellation size. It
would therefore be relatively simple to find optimal
\emph{rectangular} \gls{apsk} constellations for $M > 16$ because in
this case not many choices exist. Obviously, these constellations
might then be far away from the performance of the true optimal
constellation, but they might still offer a significant performance
gain over rectangular QAM constellations in the presence of
\gls{nlpn}, with the advantage that a constructive labeling method is
readily available.

\section{Conclusion}
\label{sec:conclusion}

In this paper, we optimized \gls{apsk} constellations for a
fiber-optical channel model without dispersion. It was shown how to
derive the \gls{pdf} of the postcompensated observation assuming a
\gls{ts} detection scheme. The \gls{pdf} was used to gain insight into
the performance behavior with respect to optimal detection and to
calculate the \gls{bep}. Optimal \gls{apsk} constellations under
\gls{ts} detection have been presented. For $M=16$ constellation
points, significant performance improvements in terms of \gls{sep} can
be achieved by choosing an optimized \gls{apsk} constellation compared
to a baseline 16-QAM constellation. For very high input power, we
showed that sacrificing points or constellation rings may become
beneficial. Finally, the binary labeling problem was discussed and a
constructive labeling method was presented, which is applicable to
rectangular \gls{apsk} constellations. An important topic for future
work would be the investigation of the influence of fiber chromatic
dispersion and nonlinearities on the optimal signal set.

\section*{Acknowledgment}

The authors would like to thank L. Beygi and F.  Brännström for many
helpful discussions.

\end{document}